\documentclass[12pt]{revtex4}
\usepackage{epsfig}
\usepackage{url,braket,amssymb}
\usepackage{xcolor,soul}

\newcommand\diag{\mathop{\mathrm{diag}}}
\newcommand\Tr{\mathop{\mathrm{Trace}}}

\newcommand\tB{\tilde B}
\newcommand\hB{\hat B}
\newcommand\hb{\hat b}
\newcommand\hP{\hat P}
\newcommand\EE{\mathbf{E}}
\newcommand\smin{\sigma_{\mathrm{min}}}
\newcommand\smax{\sigma_{\mathrm{max}}}
\begin{document}
\title{Noninvasively improving the orbit-response matrix
  while continuously correcting the orbit}
\author{Ingvar Ziemann}
\affiliation{KTH, Royal Institute of Technology, Stockholm}
\author{Volker Ziemann}
\affiliation{Uppsala University}
\date{\today}
\begin{abstract}\noindent
  Based on continuously recorded beam positions and corrector excitations from,
  for example, a closed-orbit feedback system we describe an algorithm that
  continuously updates an estimate of the orbit response matrix. The speed of
  convergence can be increased by adding very small perturbations, so-called
  dither, to the corrector excitations. Estimates for the rate of convergence
  and the asymptotically achievable accuracies are provided.
\end{abstract}
\maketitle
%
%
\section{Introduction}
The orbit-response matrix relates changes of the dipole corrector
magnets to orbit changes that are observed on the beam position monitor
system. It is of paramount importance for maintaining stable beam positions
in storage rings, which is typically accomplished by ``slow'' orbit correction
systems~\cite{KOUTCHOUK,MIZI,HUANG,VZCOR} and ``fast'' feedback systems~\cite{REHM,MIZRA}.
They either use a response matrix generated from a computer model of the
accelerator or a measured matrix found by varying one corrector at a time
and observing the ensuing changes with the beam position monitor (BPM) system.
\par
As a matter of fact, comparing the measured matrix with a matrix derived
from a computer model, as discussed in~\cite{CORBETT, SAFRANEK,DEBUGGING},
makes it possible to track down deficient hardware, such as incorrectly
calibrated power supplies or scale errors on position monitors. Usually, the
response matrix is measured in dedicated shifts, labeled ``machine development,''
  where the excitation of one corrector after the other is varied and the resulting
  changes of the positions on the orbit monitor system are recorded, which is
  commonly referred to as ``open loop'' measurements. In this report, we discuss
  an algorithm that complements the existing methods. It requires no dedicated
  beam time and slowly improves an estimate of the response matrix quasi for free
  by using information from the ``closed loop'' orbit feedback system.
  The procedure, based on a recursive least-squares algorithm~\cite{SYSINF,LJUNG},
  is completely non-invasive and can run while operating the
  accelerator in production mode---producing luminosity in a collider,
  or photons in a light source. It has the remarkable property that the
  error bars asymptotically approach zero as the estimated response matrix
  approaches the ``real'' response matrix. The algorithm is,  
  however, slow, because it ``learns from noise'' but might
  nevertheless prove useful to continuously improve the response matrix at
  times normally not accessible for machine improvement.
  This opens the possibility to track down
  very slow changes of hardware parameters when post-processing the response
  matrix with, for example, LOCO~\cite{SAFRANEK}.
\par
  This report is organized as follows:
  in the next section we develop the algorithm, followed by Section~\ref{sec:sim},
  where we introduce a simple model storage ring used to illustrate it. In
  Section~\ref{sec:dith} we introduce dithering as a way to speed up the algorithm,
  before we explore its convergence properties, both during the early stages
  in Section~\ref{sec:convergence}, and in the asymptotic regime in Section~\ref{sec:conv}.
  Before concluding, we address a number of technical issues and extensions to
  the algorithm in Section~\ref{sec:tech}.
\section{The algorithm}
The response matrix $B$ with matrix elements $B^{ij}$ relates the change
in excitation $u^j$ of steering magnet $j$ with $1\leq j\leq m$ to a
change of the beam position $x^i$ with $1\leq i\leq n$ on monitor $i$.
Here superscripts denote different monitors and correctors. We will use
the notation from quantum mechanics with bra states denoting column
vectors and ket states denoting row vectors, which will prove convenient
later on. We thus collectively denote the values of all $n$ BPM by $\ket{x}$
and the $m$ correctors by $\ket{u}$. Correcting the orbit then means to
to add a perturbation $B\ket{u}$ to the orbit $\ket{x}$ that minimizes
the residual orbit $\ket{\tilde x}$ after correction. It is given by
\begin{equation}\label{eq:respmat}
  \ket{\tilde x} = \ket{x}+B \ket{u} + \ket{w}\ ,
\end{equation}
where $\ket{w}$ describes noise in the system, for example, due to ground
motion or BPM noise. When correcting the orbit, we have to find corrector
excitations $\ket{u}$ that minimize $\braket{\tilde x| \tilde x}$. One
problem is that we do not have complete knowledge of the system matrix $B$.
All we do know is a more or less accurate estimate $\tB$ that was previously
derived from a computer model or from measurements and use that when
correcting the orbit.
\par
Assuming that the position monitors report values $\ket{x}$ and furthermore
assuming that the desired orbit is centered around zero, allows us to
calculate the desired corrector excitations $\ket{u}$ from inverting $\tB$,
the approximation of $B$ from Equation~\ref{eq:respmat}. If $\tB$ is square
($n=m$) and invertible this is just the matrix-inverse $-\tB^{-1}$, where
the minus sign ensures that the effect of the correctors cancels the observed
orbit. If $\tB$ is over-determined ($n > m$) this is accomplished by the
Moore-Penrose pseudo inverse $-(\tB^{\top}\tB)^{-1}\tB^{\top},$ which follows
from minimizing $\left(\ket{x}+\tB\ket{u}\right)^{\top}\left(\ket{x}+\tB\ket{u}\right)$
with respect to $\ket{u}$. If $\tB$ is under-determined ($n<m$) it can
be inverted using singular value decomposition. In general, we denote the
linear dependence of the corrector excitations on the observed orbit
$\ket{x}$ by the ``correction'' matrix $K$, such that $\ket{u} = -K \ket{x}$.
\par
Our task is now to extract information from repeatedly correcting the orbit
and correlating the orbit change with the used corrector changes $\ket{u}$.
To this end we note that the noise $\ket{w}$ and the mismatch of the ``real''
accelerator model $B$ and $\tB$ from which $K$ is derived, causes
the correction to be imperfect. We model this dependence by the dynamical
system
\begin{equation}\label{eq:dynsys}
  \ket{x_{t+1}}=\ket{x_t}+B\ket{u_t} + \ket{w_t}
  \qquad\mathrm{with}\qquad
  \ket{u_t}=-K \ket{x_t}\ ,
\end{equation}
where the subscript $t$ denotes a discrete time step from one iteration
of the orbit correction to the next.
  We assume that the noise $\ket{w_t}$
  is Gaussian and characterized by the expectation value 
  $\mathbf{E}\{\ket{w_s}\bra{w_t}\}=\sigma_w^2C\delta_{st}$. Here $\sigma_w^2C$
  is the spatial covariance matrix, where $\sigma_w$ is the rms magnitude
  and $C$ describes correlations among different BPM. In
  Appendix~\ref{sec:appCorr} we will return to the general case, but assume
  $C$ to be a $n\times n$ unit matrix in the main text. Furthermore,
  $\delta_{st}$ is the Kronecker delta, which implies that we treat noise
  to be uncorrelated from one iteration to the next.
    Note also that the effect of power supply noise
  $\ket{\tilde v_t}$ added to $\ket{u_t}$ is equivalent to additional
  noise on the monitors with magnitude~$B\ket{\tilde v_t}.$
In Equation~\ref{eq:dynsys} we implicitly omit fast time-dependent effects,
such as latency in the power supplies or the computation chain as well as
the effect of eddy currents.
  In Section~\ref{sec:tech} we briefly discuss
how to include these effects, but in the main text all transient
effects are assumed to have settled to a new equilibrium from one
iteration to the next. Now the interpretation of Equation~\ref{eq:dynsys} is
straightforward: the system responds with the ``real'' response matrix
$B$ to a change of the corrector excitation by $\ket{u_t}$ that was calculated with the
approximative inverse $K$ and the orbit $\ket{x_t}$. At the same time,
noise enters the system through $\ket{w_t}$, such that the residual orbit
$\ket{x_{t+1}}$ after the correction is not necessarily equal to zero.
Iterating the orbit correction, which is what orbit feedback systems
essentially do, can now be modeled by iterating the system described by
Equation~\ref{eq:dynsys}.
\par
In order to find an estimate $\hB$ of the system matrix $B$ one
row---corresponding to a particular BPM $i$---at a time,
  we construct linear systems of equations for each time step and solve
  the resulting sequence of equations with a recursive least-squares
  algorithm~\cite{SYSINF,LJUNG}. To set up the equations, for the time being,
we ignore the noise $\ket{w_t}$ and formulate Equation~\ref{eq:dynsys}
for this BPM as a constraint for $\hB$. Writing the constraints over
consecutive readings $x^i_s$, we find
\begin{equation}
  x^i_{s+1} - x^i_s
  =\left(\hB^{i1},\dots,\hB^{im}\right)
  \left(\begin{array}{c} u^1_s, \\ \vdots, \\ u^m_s\end{array}\right)
  =\left(
    \begin{array}{ccc}
           u^1_s & \dots & u^m_s \\
    \end{array}
  \right)
  \left(\begin{array}{c} \hB^{i1} \\ \vdots \\ \hB^{im}\end{array}\right)\ ,              
\end{equation}
where the second equality follows from exchanging the order of writing the
scalar product of row $i$ of $\hB$ and corrector excitations $u^j_s$. In the next
step we assemble multiple copies of this equation from different times $1\leq s\leq T$
in the form of a matrix
\begin{equation}\label{eq:defU}
  \left(\begin{array}{c} x^i_{2} - x^i_1\\ \vdots \\ x^i_{T+1} - x^i_T\end{array}\right)         
  =\left(
    \begin{array}{ccc}
           u^1_1 & \dots & u^m_1 \\
           & \vdots & \\
           u^1_T & \dots & u^m_T \\      
    \end{array}
  \right)
  \left(\begin{array}{c} \hB^{i1} \\ \vdots \\ \hB^{im}\end{array}\right)
  = U_T
  \left(\begin{array}{c} \hB^{i1} \\ \vdots \\ \hB^{im}\end{array}\right)
  = U_T\ket{\hB^{i:}}
\end{equation}
and denote the matrix containing the corrector excitations $u^j_s$ by $U_T$,
which thus contains the excitations of all correctors stacked one by one
on top of the other. Likewise, the vector on the left-hand side contains
the orbit differences that each of the steering magnet excitations causes.
If we now record BPM positions and
corresponding corrector excitations for a long time $T$, the system of
equations in Equation~\ref{eq:defU} is vastly over-determined, provided
that the noise really affects all possible degrees of freedom of the system,
  which implies that in general the covariance matrix $C$
  must have full rank. Since we assume $C$ to be the unit matrix, this
  is the case and we can solve Equation~\ref{eq:defU}
in the least-squares sense with the pseudo-inverse mentioned above. We find
\begin{equation}\label{eq:defB}
  \ket{\hB^{i:}_T}=
  \left(\begin{array}{c} \hB^{i1}_T \\ \vdots \\ \hB^{im}_T\end{array}\right)
  =\left(U_T^{\top}U_T\right)^{-1}U_T^{\top}
  \left(\begin{array}{c} x^i_{2} - x^i_1\\ \vdots \\ x^i_{T+1} - x^i_T\end{array}\right)\ ,      
\end{equation}
which provides an estimate for row $i$ of the matrix $\hB^{i:}_T$ after $T$ iterations
of the orbit corrections. Repeating this procedure for all BPMs provides us with
an estimate for the complete system matrix $\hB_T$.
\par
In passing, we point out that
Equation~\ref{eq:defB} describes a linear map from the vector with the position
differences on the right-hand side onto the vector with row $i$ of $\hB_T$, which
allows us to calculate an empirical (data-driven) covariance matrix of the $\hB_T$
from the covariance matrix of the position difference, which is $2\sigma^2_w$ times the
unit matrix. The error bars $\sigma(\hB)$ of the fitted $\hB_T$ are therefore approximately
given by the square root of the diagonal elements of $2\sigma_w^2\left(U_T^{\top}U_T\right)^{-1}$,
which will prove useful later on.
\par
Calculating the pseudo-inverse of $U_T$ for more and more iterations becomes
numerically very expensive. There is, however, an elegant way of iteratively
updating the pseudo-inverse using the Sherman-Morrison~\cite{SHEMO} formula.
It is based on updating $P_T=\left(U_T^{\top}U_T\right)^{-1}$ and $\hB_T$ as the matrix
$U_T$ grows one row at a time by adding the row vector $\bra{u_{T+1}}
=(u^1_{T+1}, \dots,  u^m_{T+1})$ to it.
This entails that we can write $P_{T+1}^{-1}=P_T^{-1}+\ket{u_{T+1}}\bra{u_{T+1}}$.
In Appendix~\ref{sec:appSM} we show that its inverse is given by
\begin{equation}\label{eq:upP}
P_{T+1} = P_T - \frac{P_T\ket{u_{T+1}}\bra{u_{T+1}}P_T}{1+\braket{u_{T+1}|P_T|u_{T+1}}}\ .
\end{equation}
With $P_{T+1}=\left(U_{T+1}^{\top}U_{T+1}\right)^{-1}$ known, we can calculate an updated
approximation~\cite{SYSINF2} of the response matrix $\hB_{T+1}$ from
\begin{equation}\label{eq:upB}
  \hB_{T+1}=\hB_T+\frac{\left(\ket{x_{T+2}}-\ket{x_{T+1}}-\hB_T\ket{u_{T+1}}\right)\bra{u_{T+1}}P_T}
  {1+\braket{u_{T+1}|P_T|u_{T+1}}}\ .
\end{equation}
We refer to Appendix~\ref{sec:appCC} and~\cite{SYSINF2} for the derivation. Note that
in Equations~\ref{eq:upP} and~\ref{eq:upB} the right-hand sides only depend on $P_T$
and $\hB_T$ from the previous iteration, the new corrector excitations $\ket{u_{T+1}}$,
and the most recent change in the orbit $\ket{x_{T+2}}-\ket{x_{T+1}}$. These two equations
now allow us to continuously update the response matrix while correcting the orbit
with~$K$. All we do here is correlating the change of the orbit $\ket{x_{T+2}}-\ket{x_{T+1}}$
with the corrector excitations $\ket{u_{T+1}}$ that cause this change and then update
our approximation of $\hB$ in the process.
\par
With the basic algorithm worked out, we simulate its performance in the following
section.
\section{Simulation}
\label{sec:sim}
%
\begin{figure}[tb]
\begin{center}
\includegraphics[width=0.47\textwidth]{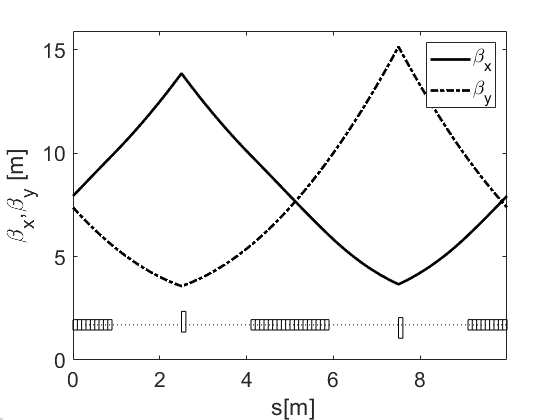}
\includegraphics[width=0.47\textwidth]{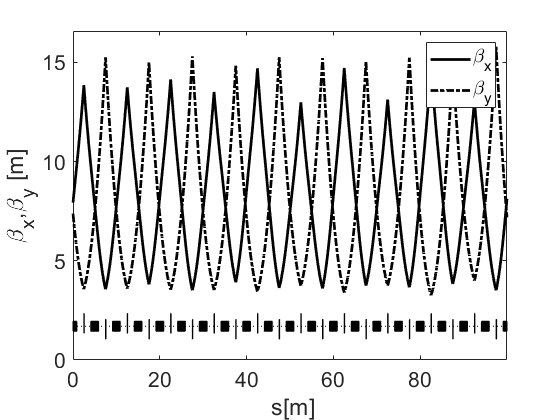}
\end{center}
\caption{\label{fig:betacell}Left: the beta functions of one cell. Right:
  the beta functions of the ring where the focal lengths of all quadrupoles
  are randomly perturbed by 5\,\% which causes a moderate beating.}
\end{figure}
In order to test the algorithm we prepared response matrices for a small ring
consisting of ten FODO cells each having phase advances of $\mu_x/2\pi=0.228$ and
$\mu_y/2\pi=0.238$ in the horizontal and vertical plane respectively. The tunes
of the ring therefore are $Q_x=2.28$ and $Q_y=2.38.$ Moreover, there are two
18-degree sector dipole magnets in each cell. The beta functions of one cell
are shown on the left-hand side in Figure~\ref{fig:betacell}. We place a corrector
and a BPM at the same location as the (thin-lens) focusing quadrupole, which
then accounts for ten correctors and ten BPM, each. In order to keep the simulation
transparent, we calculate the response matrix $B$ between these correctors
and BPM in the horizontal plane only. Most of the simulations are done for equal
numbers of correctors and monitors; we address other cases in Section~\ref{sec:tech}.
The response matrix derived from this unperturbed ring is the
``ideal'' response matrix $\tB$ that we use to derive the correction matrix
$K=\left(\tB^{\top} \tB\right)^{-1}\tB^{\top}$ to correct the orbit. In order to simplify
the theoretical analysis in Section~\ref{sec:conv}, in the remainder of this report
we confine ourself to a constant correction matrix $\tilde B$. If, instead we
were to use the constantly updating $\hB_T$ for the correction, the algorithm
would be adaptive. In order to determine the ``real'' response matrix $B$, we
randomly vary the focal lengths of the quadrupoles with a rms of 5\,\% and
re-calculate the response matrix $B$ for the perturbed ring.  We take notice
that the rms magnitude of the response coefficients is 6.6\,m/rad. In order to
quantify the estimation error $b_T=\hB_T-B$ after $T$ iterations, we introduce
the rms value of $b_T$, calculated over all matrix elements as {\em discrepancy}
$|b_T|_{rms}$. It can also be calculated from
\begin{equation}
  |b_T|_{rms}=\sqrt{\frac{\Tr\left((\hB_T-B)^{\top}(\hB_T-B)\right)}{nm}}\ .
\end{equation}
Evaluating the initial value $|b_0|_{rms}$ for our model storage ring, we find that
it is approximately $0.3\,$m/rad which accounts for a 5\,\% rms deviation of the
response matrix coefficients. The simulations are based on Matlab scripts that
use beam optics functions from~\cite{VZACC}. The code illustrating one iteration
of the algorithm is reproduced and commented in Appendix~\ref{sec:appCode}.
\par
\begin{figure}[tb]
\begin{center}
\includegraphics[width=0.85\textwidth]{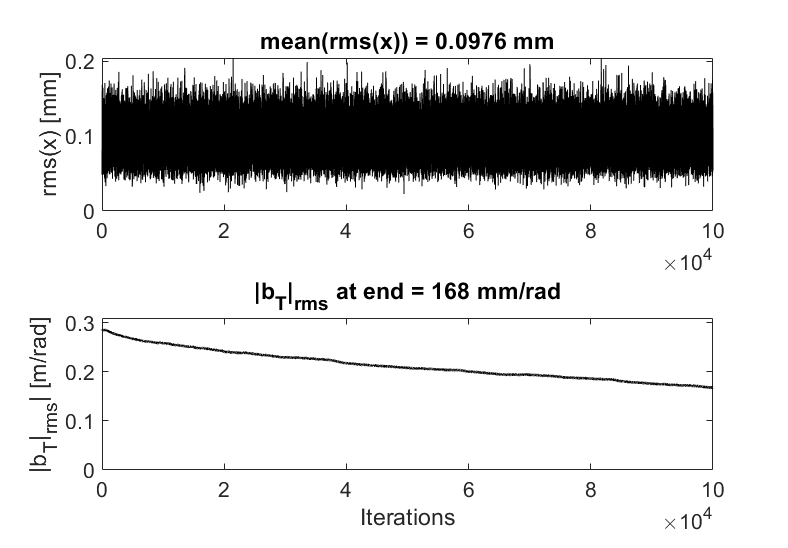}
\end{center}
\caption{\label{fig:sig010}The rms orbit $\sqrt{\braket{x|x}}$ (top) and
  the rms of all matrix elements of $B-\hat B$ (bottom) for $10^5$ iterations.
  The noise level was chosen $\sigma_w=0.1\,$mm. The upper plot reproduces the
  noise level and the lower graph shows that the response-matrix estimate $\hat B$
  slowly converges towards $B$.}
\end{figure}
Running the simulation for $10^5$ iterations, which takes a few seconds
on a desktop computer, produces Figure~\ref{fig:sig010}, which shows the
evolution of the rms orbit $\sigma_x^2=\sqrt{\braket{x|x}}$ and the
discrepancy~$|b_T|_{rms}$
between the ``real'' and the estimated response matrix. We initialized the
estimate for $\hB_0$ with the response matrix for the ring without quadrupole
gradient errors and $P_0$ with the unit matrix. The rms amplitude of the
noise $\sigma_w$ was chosen to be 0.1\,mm.
Note that the upper plot, which shows the rms orbit $\sigma_x$ for the duration of
the simulation clearly verifies this; the mean is close to 0.1\,mm. At the
same time, the discrepancy~$|b_T|_{rms}$, shown on the lower plot, is approximately
halved to a final value $|b_f|_{rms}=0.168\,$m/rad, which shows that the
algorithm works.
\par
\begin{figure}[tb]
\begin{center}
\includegraphics[width=0.65\textwidth]{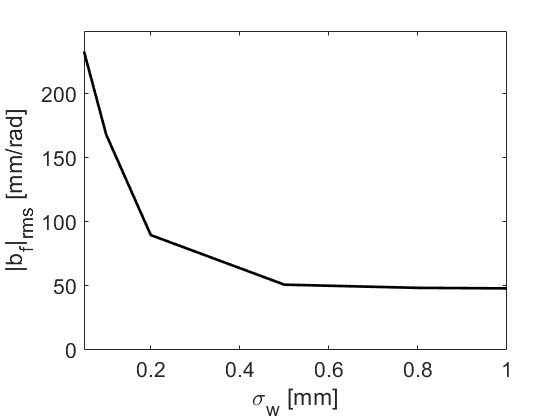}
\end{center}
\caption{\label{fig:scalsig}The rms value of all matrix elements of $B-\hat B$
  after $10^5$ iterations as a function of the noise level $\sigma_w$.}
\end{figure}

Repeating the same simulation (always for $10^5$ iterations) for different
values of $\sigma_w$ and recording the final discrepancy $|b_f|_{rms}$ produces
the plot shown in Figure~\ref{fig:scalsig}. Here we find that increasing noise
levels are beneficial for the rate of convergence, up to about $\sigma_w=0.5\,$mm,
where the induced changes in $b$ during one iteration become comparable to the
magnitude of $b$. We need to stress that the plotted values are those reached
after $10^5$ iterations. They are not the asymptotic levels.
\par
\par
The algorithm is rather stable. We ran simulations where we initialized $\hB$
with random matrices or other made-up starting guesses. The algorithm, after
an initial transient period, always converged towards the ``real'' matrix~$B$.
\par
We point out that the convergence depends on the noise level, where more noise
moves the correctors around more and actually improves the convergence, but the
rate is still rather slow, on the order of several $10^5$ iterations, which would
correspond to about three hours real time, provided that the feedback operates at
an update rate of 10 per second. Moreover, the asymptotically achievable discrepancy
is of considerable interest. We will address these topics below after having
introduced the effect of additional corrector perturbations. 
\section{Dithering}
\label{sec:dith}
%
\begin{figure}[tb]
\begin{center}
\includegraphics[width=0.47\textwidth]{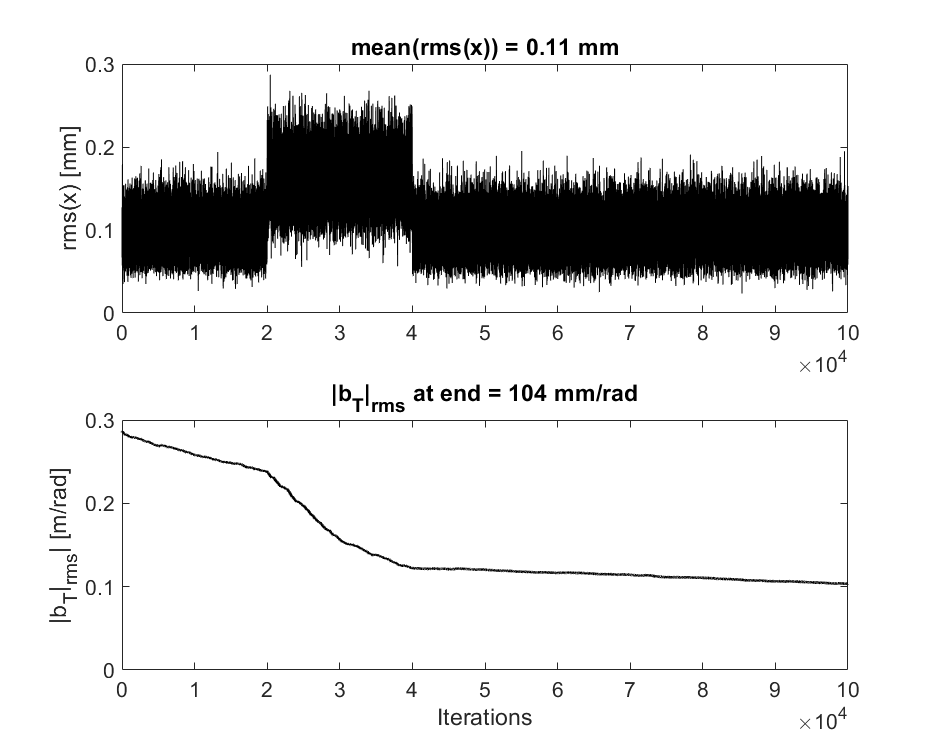}
\includegraphics[width=0.47\textwidth]{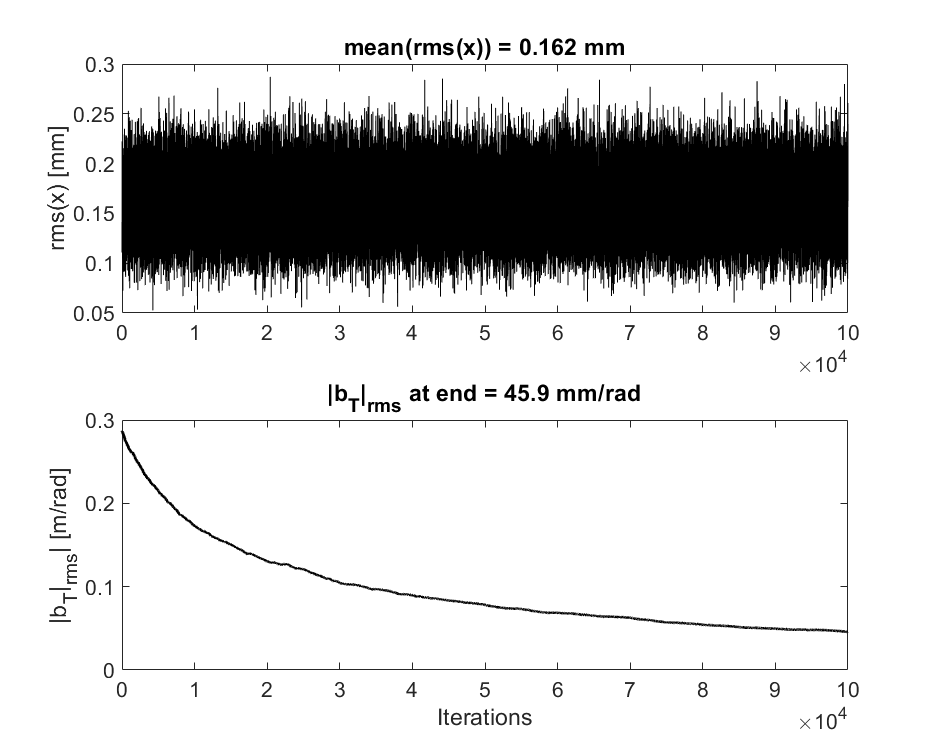}
\end{center}
\caption{\label{fig:dither02}Simulations based on parameters used for Figure~\ref{fig:sig010}.
  Left: round-robin dithering with $20\,\mu$rad active from 20000 to 40000 iterations,
  which temporarily increases the rms orbit $\sigma_x$ but significantly helps to reduce
  the discrepancy $|b_{rms}|$. Right: dithering with the same parameters is active all the
  time, which increases $\sigma_x$ all the time, but reduces $|b_{rms}|$ even further.}
\end{figure}
Varying the corrector excitations one at a time, either systematically or
sinusoidally~\cite{DIAMOND,ACLOCO,ALBA}, in order to determine the response matrix
is used in practically all accelerators.
Moreover, continuously varying correctors very little such that the detrimental effect
on the orbit is negligible, so-called dithering, was successfully used~\cite{MCROSS,PEP2,KEKB}
to optimize the performance of a number of accelerators. We implement dithering in our
simulation by adding a perturbing vector $\ket{z_t}$ to $\ket{u_t}$ when correcting the orbit in
Equation~\ref{eq:dynsys}, which therefore becomes $\ket{u_t}=-K \ket{x_t} + \ket{z_t}$.
%
The rest of the simulation remains unaffected; any changes of $\ket{z_t}$ and consequently
of $\ket{u_t}$ are consistently accounted for in the updates of $P_T$ and $\hB_T$ in
Equations~\ref{eq:upP} and~\ref{eq:upB}.
\par
In the simulations, shown in Figure~\ref{fig:dither02}, we chose to add $20\,\mu$rad
to the excitation $z^k$ to one corrector~$k$ at a time in a round-robin fashion and record
the rms orbit and the discrepancy for $10^5$ iterations. The plot on the left-hand
side shows the simulation where the dithering was turned on between 20000 and 40000
iterations. We clearly see that the rms orbit increases from 0.10 to 0.16\,mm during this
period, which is consistent with expectations, because the rms value of the $B$ of 6.6\,m/rad
and $20\,\mu$rad additional excitation results in an additional rms orbit variation
of $0.13\,$mm, which, added in quadrature to $\sigma_w=0.1\,$mm, gives about 0.16\,mm.
We also observe on the lower plot that the discrepancy $|b_T|_{rms}$ is significantly
reduced and conclude that temporarily adding dithering helps to improve our knowledge
of the response matrix. Note that no additional processing of the data is necessary.
The algorithm learns whenever it gets the chance to observe some variation, never mind
the source of the perturbation. Remarkably, a slammed door might be beneficial for
something. In the simulation shown on the right-hand side in Figure~\ref{fig:dither02},
we keep the $20\,\mu$rad round-robin dithering on permanently and observe that the
rms orbit is 0.16\,mm throughout the simulation, while the discrepancy~$|b_T|_{rms}$
is reduced sevenfold. Again, no special processing is required.
\par
\begin{figure}[tb]
\begin{center}
\includegraphics[width=0.47\textwidth]{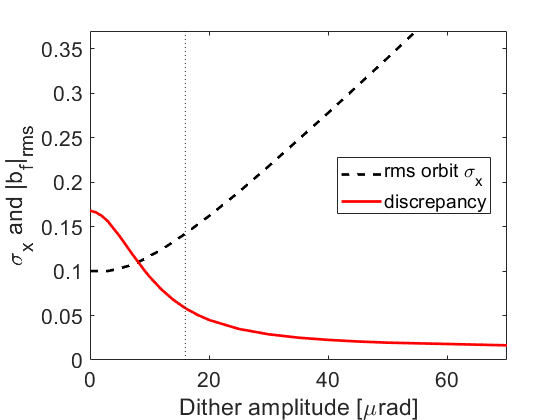}
\includegraphics[width=0.47\textwidth]{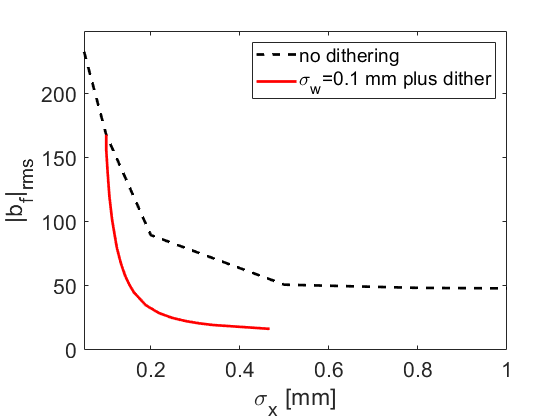}
\end{center}
\caption{\label{fig:scadit}Left: the rms orbit $\sigma_x$ (dashed black) and the
  discrepancy $|b_{rms}|$ (solid red) as a function of the dither amplitude,
  which allows us to assess the trade-off between spoiling the orbit and
  learning the response matrix. Right: the solid red line shows $|b_f|_{rms}$ plotted
  versus $\sigma_x$. The black dashed line shows the effect of purely random
  variations, already shown in Figure~\ref{fig:scalsig}, for comparison.}
\end{figure}
The left-hand plot in Figure~\ref{fig:scadit} illustrates the effect of dither
amplitude, shown on the horizontal axis, on the rms orbit (dashed black) and on
the discrepancy (solid red). We clearly observe that the increasing dither amplitude
increases the rms orbit $\sigma_x$, but at the same time, helps to reduce the
discrepancy $|b_f|_{rms}$. Closer inspection shows that a dither amplitude of
$16\,\mu$rad contributes to $\sigma_x$ with the same magnitude as normal noise
level $\sigma_w$. This causes $\sigma_x$ to increase by 40\,\%. At the same time,
$|b_f|_{rms}$ is reduced by $1/3$ from 0.168\,m/rad to 0.056\,m/rad.
This configuration is indicated by the vertical dotted line in Figure~\ref{fig:scadit}.
\par
The right-hand plot in Figure~\ref{fig:scadit} shows the data from the left-hand
plot, but now plotting the discrepancy $|b_f|_{rms}$ versus the the rms orbit $\sigma_x$
(solid red) and compares it to the data from Figure~\ref{fig:scalsig} (dashed black).
Unsurprisingly, increasing $\sigma_x$ by dithering reduces $|b_f|_{rms}$ more efficiently
than just increasing the natural noise level $\sigma_w$.
\section{Convergence}
\label{sec:convergence}
%
\begin{figure}[tb]
\begin{center}
\includegraphics[width=0.8\textwidth]{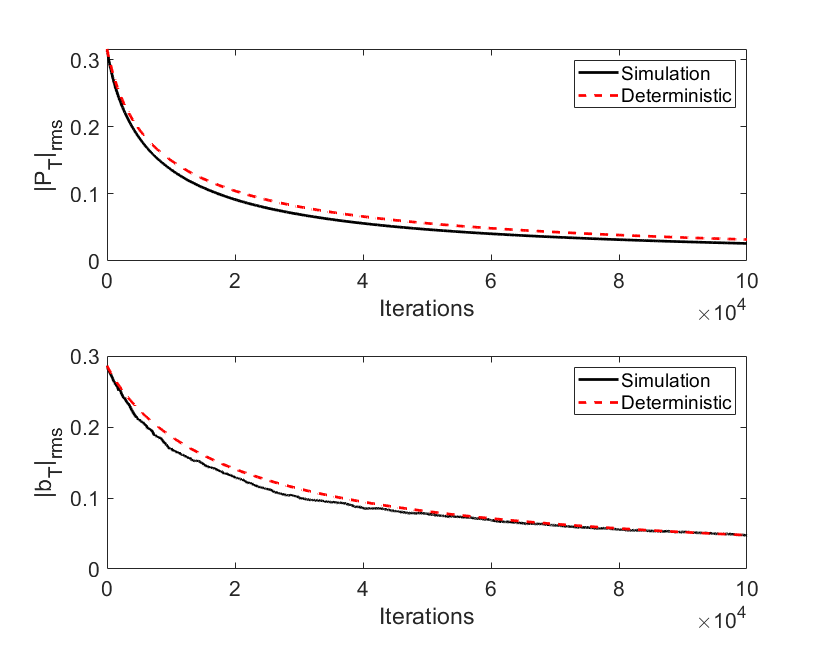}
\end{center}
\caption{\label{fig:simdet}The upper panel shows the evolution of $|P_T|_2$
  (solid black) and $|\hP_T|_2$ (dashed red) for a configuration with
  $\sigma_w=0.1\,$mm and $z=20\,\mu$rad. The lower panel shows the corresponding
  evolution of $b_T$ (solid black) and $\hb_T$ (dashed red).}
\end{figure}
A matter of practical interest are the time scales, given by the number of
iterations, before we observe some improvement of the response matrix.
  We point out that the results developed in the following sections
  apply to all systems described by
  Equation~2, which includes
  rings with transverse coupling and correction matrices $K$ that use
  elaborate regularization schemes. The simulations, which are based
  on correctors and monitors in a single transverse plane, are only
  used to illustrate the general results.
Let us start by analyzing
the initial behavior of the discrepancy and approximate Equation~\ref{eq:upP}
by replacing $\ket{u_{T+1}}\bra{u_{T+1}}$ by its expectation value
$\EE\{\ket{u_{T+1}}\bra{u_{T+1}}\}$, which asymptotically becomes independent of $T$.
We therefore use $\EE\{\ket{u_{T}}\bra{u_{T}}\}$ instead, which depends on $\ket{x_T}$
via $\ket{u_T}=-K\ket{x_T}$ and calculate
\begin{eqnarray}
  \ket{x_T}&=&(1-BK)\ket{x_{T-1}}+\ket{w_{T-1}}\nonumber\\
           &=&(1-BK)^T\ket{x_0} +\sum_{s=0}^{T-1}(1-BK)^s\ket{w_{T-s-1}}\ ,
\end{eqnarray}
where the second equality results from iterating the first equality. Since the spectral
radius $\rho(\Lambda)$ with $\Lambda=1-BK$ is much less than unity, the influence of
the initial $\ket{x_0}$ ``dies out'' for large $T$ and we can omit the first term
from the sum. Inserting $\ket{x_T}$ in $\ket{u_T}=-K\ket{x_T}$, we obtain 
\begin{eqnarray}\label{eq:EEuu}
  \EE\{\ket{u_T}\bra{u_T}\}
  &=& \EE\left\{K\left[ \sum_{s=0}^{T-1}\Lambda^s\ket{w_{T-s-1}}\right]
      \left[\sum_{r=0}^{T-1}\bra{w_{T-r-1}}\left(\Lambda^{\top}\right)^r\right] K^{\top}\right\} +o(1)
      \nonumber\\
  &=& K\sum_{s=0}^{T-1}\sum_{r=0}^{T-1}\Lambda^s
      \EE\left\{\ket{w_{T-s-1}}\bra{w_{T-r-1}}\right\}\left(\Lambda^{\top}\right)^r K^{\top}+o(1)\\
  &=&  \sigma_w^2K\sum_{s=0}^{T-1} \left(\Lambda\Lambda^{\top}\right)^sK^{\top}+o(1)\ ,\nonumber
\end{eqnarray}
where we used that the expectation value of the Gaussian noise is
$\mathbf{E}\{\ket{w_s}\bra{w_t}\}=\sigma_w^2\delta_{st}\mathbf{1}$.
Moreover, $o(1)$ denotes a quantity that vanishes in the limit of large $T$.
The smallness of $\rho(\Lambda)$ implies that only the term with $s=0$ in the sum in
Equation~\ref{eq:EEuu} contributes and we have $\EE\left\{\ket{u_{T}}\bra{u_{T}}\right\}
\approx\sigma_w^2K K^{\top}$, which is indeed independent of $T$. We include round-robin
dithering with amplitude $z$ through
the $m$ correctors by adding a term $\frac{z^2}{m}\mathbf{1}$, because dithering 
is uncorrelated to the noise and after $m$ iterations dithering contributes a unit
matrix. We thus just ``spread out'' this unit matrix to the individual iterations
when diving by $m$. We therefore introduce 
\begin{equation}\label{eq:Q}
Q=\sigma_w^2K K^{\top} + \frac{z^2}{m}\mathbf{1}
\end{equation}
to represent the average effect of the orbit correction and dithering when
updating the ``averaged'' $\hP_T$ in Equation~\ref{eq:upP}, which then reads
\begin{equation}\label{eq:upPQ}
  \hP_{T+1}=\hP_T-\frac{\hP_T Q \hP_T}{1+\Tr\left(Q\hP_T\right)}\ .
\end{equation}
Note that Equation~\ref{eq:upPQ} is a deterministic equation that describes the
averaged updating of $\hP_T$. In the simulation we update $\hP_T$ in parallel
to its ``stochastic brethren'' $P_T$ and find that they are extremely close,
both with and without dithering. The upper panel in Figure~\ref{fig:simdet}
shows an example with $\sigma_w=0.1\,$mm and $z=20\,\mu$rad, which corresponds to
the configuration also displayed on the right-hand side in Figure~\ref{fig:dither02}.
The solid black curve is produced by a numerical simulation with simulated random
noise and the dashed red curve shows the result of the deterministic simulation,
based on Equations~\ref{eq:Q} and~\ref{eq:upPQ}. 
\par
\begin{figure}[tb]
\begin{center}
\includegraphics[width=0.6\textwidth]{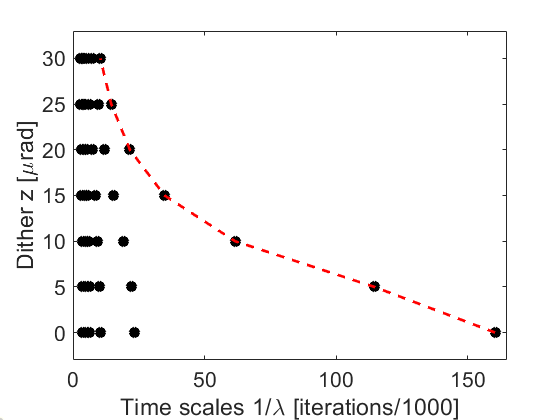}
\end{center}
\caption{\label{fig:ts}Time scales of the convergence, determined from the inverse
  eigenvalues of $Q$ for dithering amplitudes $z$ from $0$ to $30\,$mrad.
  In all cases we use $\sigma_w=0.1\,$mm. The dashed red line shows the time scale
  $1/\lambda_{\mathrm{min}}$ with $\lambda_{\mathrm{min}}=\sigma_w^2\smin\left[K K^{\top}\right]+z^2/m$
  corresponding to the smallest eigenvalue of~$Q$.}
\end{figure}
We point out that $Q$ is the only parameter in the dynamics described by
Equation~\ref{eq:upPQ}. In order to simplify the analysis somewhat, we neglect
the trace in the denominator, which is practically always much smaller than $1$,
which results in $1+\Tr\left(Q\hP_T\right)\approx 1$ and allows us to write
the equation as $\hP_{T+1}=\hP_T-\hP_T Q \hP_T$. Moreover, $Q$ is symmetric by
construction and we can choose a coordinate system in which $Q$ is diagonal
with eigenvalues $\lambda_j=\sigma_w^2\sigma_j+z^2/m$, where
$\sigma_j$ are the eigenvalues of $K K^{\top}$, such that $Q=O D O^{\top}$ with
$D=\diag(\lambda_1,\dots,\lambda_m)$ and an orthogonal matrix $O$. Also the
starting guess for $P_0$ is the unit matrix and is diagonal, such that
Equation~\ref{eq:upPQ} can be written
as $m$ independent equations for each of the $m$ diagonal elements $x_{j,T}$ of
$\hP_T$. Each eigenvalue thus corresponds to one mode that describes the
dynamics of the convergence process. In the following, we consider one mode at
a time and omit a second index $j=1\,\dots,m$ from $x$ and $\lambda$ to
make the equations easier to read. We therefore obtain $x_{T+1}=x_T-\lambda x_T^2$
or its continuous approximation $dx_T/dT=-\lambda x_T^2$ for each mode. This
equation has the solution
\begin{equation}\label{eq:x}
x_T=\frac{x_0}{1+x_0\lambda T}\ .
\end{equation}
Numerically $x_0$ has the value of unity, because $\hP_0$ is the unit matrix,
but we leave it in place to keep track of the units of $x_0$ which are 1/mrad$^2$.
We thus find that the inverse eigenvalues $1/\lambda$ of the matrix $Q$ determine
the time scales of the convergence of the process. Note, however, that the time
dependence is inversely proportional to $T$, rather than exponential, and is
therefore slow.
\par
Figure~\ref{fig:ts} shows the time scales $1/x_0\lambda_j$ with $j=1,\dots,m$ for
dither amplitudes $z$ between 0 and $30\,$mrad, while $\sigma_w$ is always $0.1\,$mm.
The dashed red line shows the time scale of the slowest mode and is given by the
smallest eigenvalue $\lambda_{\mathrm{min}}=\sigma_w^2\smin\left[K K^{\top}\right]+z^2/m$.
Here $\smin[\cdot]$ is the smallest eigenvalue of the matrix in the argument.
The rms orbit variation $\sigma_x$ approximately doubles in this range.
We observe that there is always one very small eigenvalue, which leads to a
very long time scale. Dithering mostly helps to reduce this long time
scale from $1.6\times 10^5$ to about 20000 iterations.
At this point we remind ourselves, following the discussion from Section~2, that
$P_T=\left(U_T^{\top}U_T\right)^{-1}$ determines the error bars of $\hB_T$. Since
Equation~\ref{eq:x} implies $P_T\propto 1/T$ for $T\gtrapprox 1/x_0\lambda$ we find that
the error bars of $\hB_T$ have a $1/\sqrt{T}$ dependence.
\par
It remains to analyze the time scales of the convergence of the values of $\hB_T$
to $B$, which is described by Equation~\ref{eq:upB}. We note that $\ket{x_{T+2}}-\ket{x_{T+1}}
=B\ket{u_{T+1}}$ was caused by the corrector values $\ket{u_{T+1}}$, such that we
arrive at
\begin{equation}\label{eq:dBiter}
  \left(\hB_{T+1} -B\right) = \left(\hB_T-B\right) - \left(\hB_T-B\right)
  \frac{\ket{u_{T+1}}\bra{u_{T+1}}P_T}{1+\braket{u_{T+1}|P_T|u_{T+1}}}\ ,
\end{equation}
where we subtracted $B$ on both sides. We now replace $\ket{u_{T+1}}\bra{u_{T+1}}$
by its expectation value and therefore use $Q$ from Equation~\ref{eq:Q} to
arrive at
\begin{equation}\label{eq:biter}
  \hb_{T+1}=\hb_T\Xi_T 
\qquad\mathrm{with}\qquad
  \Xi_T=1-\frac{Q\hP_T}{1+\Tr\left(Q\hP_T \right)}\ ,
\end{equation}
where we introduced $\hb_T=\hB_T -B$ to simplify the writing. Like Equation~\ref{eq:upPQ}
before is this a deterministic equation for $\hb_T$ that we update in parallel to the
stochastic simulations that generate $b_T$. On the bottom panel in Figure~\ref{fig:simdet}
we show $|b_T|_{rms}$, the rms value of $b_T$, as a solid black line and $|\hb_T|_{rms}$ as
dashed red line for a simulation with  parameters specified in the figure caption. We
take notice that both black and red curves track one another very well, which allows
us to determine the time scales from analyzing $\Xi_T$ from
Equation~\ref{eq:biter}. As before, we use a coordinate system in which $Q$ and
$P_T$ are diagonal, ignore the denominator with the trace, and analyze one mode at a
time. If we denote the eigenvalue of $\Xi_T$ by $\xi_j$ (and omit the index j henceforth,
\begin{figure}[tb]
\begin{center}
\includegraphics[width=0.47\textwidth]{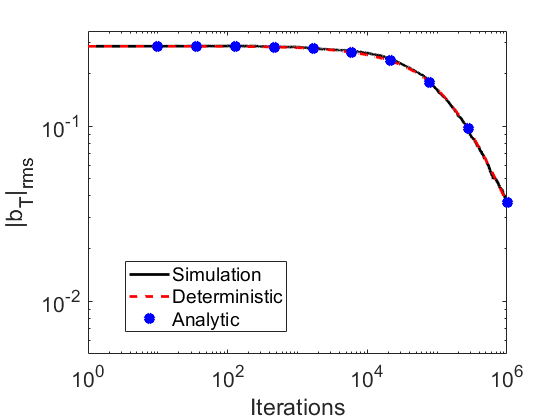}
\includegraphics[width=0.47\textwidth]{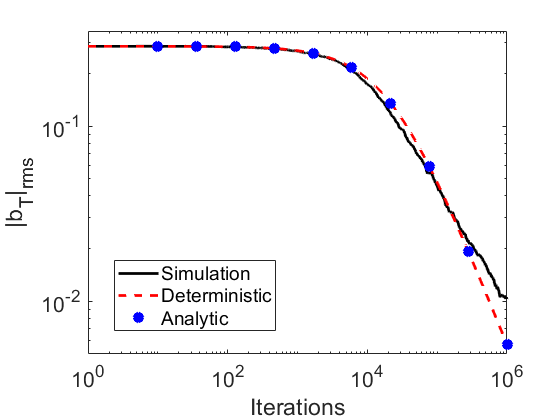}
\end{center}
\caption{\label{fig:analy}Left: the discrepancies $|b_T|_{rms}$ calculated by direct
  numerical simulation (black line), from Equation~\ref{eq:upPQ} (red dashes), and from
  Equation~\ref{eq:bG} (blue dots) for $\sigma_w=0.1\,$mm and no dither. Right: with
  $20\,\mu$rad round-robin dither added.}
\end{figure}
because we consider one mode at a time and want to use the subscript to denote the
iteration), we find
\begin{equation}
\xi_T=1-\lambda x_T = \frac{1-x_0\lambda+x_0\lambda T}{1+x_0\lambda T}\ ,
\end{equation}
where we substituted $x_T$ from Equation~\ref{eq:x}. Again the time scales are determined
by $1/\lambda$, the inverse eigenvalues of $Q$. Inspecting Equation~\ref{eq:biter},
we see that the eigenvalues $\Xi_s$ describe how the modes decrease
{\em from} one iteration $T-1$ {\em to} $T$. In order to find the total reduction {\em after}
$T$ iterations we need to multiply all the previous eigenvalues $\xi_s$ for $1\le s \le T$,
which gives us the eigenvalues $y_T$ of the product $Y_T=\prod_{s=1}^T\Xi_s$
\begin{equation}
  y_T=\prod_{s=1}^T\xi_s=\prod_{s=1}^T\frac{1-x_0\lambda+x_0\lambda s}{1+x_0\lambda s}
  = \frac{1}{1+x_0\lambda T}\ ,
\end{equation}
where the last equality is straightforward to prove by induction. Thus $Y_T$ is a diagonal
matrix with expressions $1/(1+x_0\lambda T)$ along its diagonal. If we now rewrite this equation
in non-diagonal coordinates, we obtain the matrix $G_T=O Y_T O^{\top}$ that maps the initial
$\hb_0$ to $\hb_T$ after iteration $T$ via
\begin{equation}\label{eq:bG}
  \hb_T=\hb_0G_T
  \qquad\mathrm{with}\qquad
  G_T=O\diag\left(\frac{1}{1+x_0\lambda_1T},\dots,\frac{1}{1+x_0\lambda_mT}\right) O^{\top}
\end{equation}
and $\lambda_j=\sigma_w^2\sigma_j\left[K K^{\top}\right]+z^2/m$ without iterating through
all the intermediate steps. In passing, we point out that $G_T$ behaves like a transfer
function that maps the initial $\hb_0$ to a later value $\hb_T$. Iterating with, for
example, different dither amplitudes $z$ only involves left-multiplying with different
$G_T$, each one calculated with the appropriate $z$.
\par
Figure~\ref{fig:analy} shows several discrepancies $|b_T|_{rms}$ as a function of the
iteration number using double-logarithmic scales. On the left-hand plot we use a
configuration with $\sigma_w=0.1\,$mm and no dithering. The black line shows  $|b_T|_{rms}$
from the stochastic simulation, the red line shows $|\hb_T|_{rms}$ using the deterministic
iteration, while the blue dots are calculated with the matrix $G_T$. We observe that all
three curves track one another very well. The plot on the right-hand side in Figure~\ref{fig:analy}
shows the configuration with $20\,\mu$rad dithering added, already used in Figure~\ref{fig:simdet}
with the blue dots from the analytic calculation superimposed. Again, the agreement is
rather good, though some discrepancies show up, once $|b_T|_{rms}$ becomes very small. Let
us therefore analyze this late regime more carefully.
\par
\begin{figure}[tb]
\begin{center}
\includegraphics[width=0.7\textwidth]{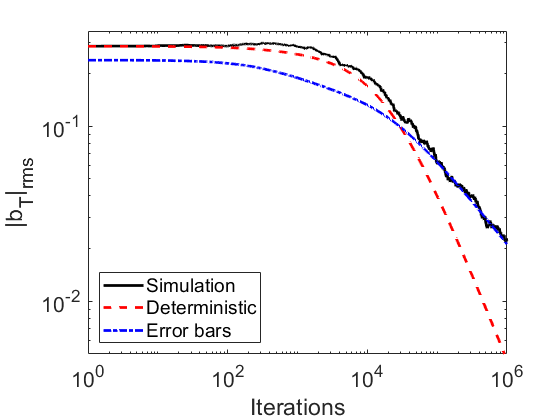}
\end{center}
\caption{\label{fig:eb}The rms value of the discrepancy $|b_T|_{rms}$ from a numerical
  simulation (solid black) and from iterating Equations~\ref{eq:upPQ} and~\ref{eq:biter}
  as well as the approximate error bars $\sigma(\hB)$ (blue dashes) for a configuration
  with $\sigma_w$ increased to $0.3\,$mm and no dither.} 
\end{figure}
From the discussion in Section~2 we know that $2\sigma_w^2P_T=2\sigma_w^2\left(U_T^{\top}U_T\right)^{-1}$
is a data-driven approximation of the covariance matrix for the matrix elements of $\hB_T$. We therefore
heuristically approximate the error bars by $\sigma(\hB)=\sqrt{2|P_T|_{rms}}\sigma_w$ and show
$|b_T|_{rms}$ for a numerical simulation (solid black) and the deterministic average (red dashes)
as well as $\sigma(\hB)$ (blue dash-dots) in Figure~\ref{fig:eb}. We observe that once $|b_T|_{rms}$
becomes smaller than $\sigma(\hB)$ the numerical simulation significantly
differs from the averaged model. In this regime the approximations, in particular, factoring
the expectation value of the product of $\hB_T-B$ and $\ket{u_{T+1}}\bra{u_{T+1}}P_T$ into separate
expectation values no longer hold. Here, the statistical fluctuations around the mean and the
$1/\sqrt{T}$ scaling of the error bars (blue dash-dots) become the dominating factor for the rate
of convergence. We therefore need to address the asymptotic regime separately, which is the
topic of the next section.
\section{Asymptotics}
\label{sec:conv}
The asymptotic regime is characterized by the discrepancy $|b_T|_{rms}$ being smaller than
the error bars, or heuristically; the signal $|b_T|_{rms}$ is inside the noise floor.
We saw in the simulations shown on the figures that even in this regime $\hB_T$ converges
towards the "real" response matrix~$B$. If we focus on cases without dithering ($z=0$), we
can explore this further by exploiting a theorem by Lai and Wei~\cite{LAIWEI}, which
states that
\begin{equation}\label{eq:laiwei}
  \vert\hB_T -B\vert_{\infty}
  = O\left(\sqrt{\frac{\log\left(\smax\left[P_T^{-1}\right]\right)}
      {\smin\left[P_T^{-1}\right]}}\right)\ ,
\end{equation}
where $\smin[\cdot]$ and $\smax[\cdot]$ denote the smallest and largest eigenvalue
of the matrix in the argument, respectively. $\vert\cdot\vert_{\infty}$
denotes the largest value of the matrix in the argument, which is always larger
than the rms value of all matrix elements that we used in the previous sections;
the two values only differ by a numerical factor of order unity. The symbol
$O(\cdot)$ denotes the leading order in $T,$ and $P_T^{-1}=U_T^{\top}U_T
=\sum_{t=1}^T\ket{u_t}\bra{u_t}$ was defined earlier. We therefore need to
determine the scaling of $\sum_{t=1}^T\ket{u_t}\bra{u_t}$ and its smallest
and largest eigenvalues with $T$. 
\par
To do so, we note that the system, defined by Equation~\ref{eq:dynsys}, can
be written as $\ket{x_{t+1}}=(1-BK)\ket{x_t} + \ket{w_t}$,
which shows that the time step $t+1$ only depends on parameters at time $t$,
which makes it a Markov chain. Moreover, if the closed-loop system is stable,
the spectral radius $\rho(\Lambda)$, with $\Lambda=1-BK$, is strictly less than
unity, which causes the process to forget all uniformly bounded initial conditions
sufficiently fast. This makes the corresponding Markov chain uniformly ergodic and
implies that the time-average and the average over the distribution function
of the noise, the expectation value $\EE\left\{\cdot\right\}$, are the same 
\begin{equation}\label{eq:ergodic}
\frac{1}{T}\sum_{t=1}^T\ket{u_t}\bra{u_t} = \EE\{\ket{u_T}\bra{u_T}\} + o(1)\ ,
\end{equation}
where, as before, $o(1)$ is an expression that vanishes in the limit of large $T$.
The right-hand side of Equation~\ref{eq:ergodic} we already calculated in
Equation~\ref{eq:EEuu} and turn to its asymptotic behavior, which is encapsulated
in the limit of $\Gamma_T=\sum_{s=0}^{T-1} \left(\Lambda\Lambda^{\top}\right)^s$ for
large $T$. First we note that
\begin{equation}\label{eq:asymG}
  \sum_{s=0}^{T-1} \left(\Lambda\Lambda^{\top}\right)^s \leq \sum_{s=0}^{\infty} \rho(\Lambda)^{2s}\mathbf{1}
  = \frac{1}{1-\rho(\Lambda)^2}\mathbf{1} < \infty
\end{equation}
is finite. Second, the existence can be proven by noting that $\Gamma_T$ is a
Cauchy sequence; $\Gamma_T-\Gamma_{T'}=o(1)$ for large $T,T'$.
We can therefore introduce $\Gamma = \lim_{T\to\infty}\Gamma_T$ and obtain
\begin{equation}\label{eq:Euu}
  \EE\{\ket{u_T}\bra{u_T}\} = \sigma_w^2K\Gamma K^{\top}+ o(1).
\end{equation}
This expression allows us to determine the smallest and largest eigenvalue
of the left-hand side 
\begin{equation}
  \smin\left[\EE\left\{\ket{u_T}\bra{u_T}\right\}\right]
  = \sigma_w^2\smin\left[K\Gamma K^{\top}\right]+ o(1)
\end{equation}
and likewise for $\smax$. We note that the smallest and largest eigenvalues of
a matrix are continuous functions of the matrix elements. This implies---as a
consequence of the continuous mapping theorem~\cite{CMT}---that limits of these
functions are preserved, even if the matrix elements depend on random variables.
We therefore obtain from Equation~\ref{eq:ergodic}
\begin{eqnarray}
  \sigma_{\mathrm{min}}\left[\sum_{t=1}^T\ket{u_t}\bra{u_t}\right]
  &=&\smin\left[T\EE\left\{\ket{u_T}\bra{u_T}\right\}+o(T)\right]\nonumber\\
  &=&\sigma_w^2 T\sigma_{\mathrm{min}} \left[K\Gamma K^{\top}\right]+ o(T)
\end{eqnarray}
and likewise for $\smax$. Here $o(T)$ denotes a quantity that increases strictly
slower with $T$ than $T$. Moreover, the convergence of the random variables on
the left-hand side towards the expectation value on the right-hand side happens
with probability~1---{\em almost surely} in the mathematical literature.
Summarily, both $\smin\left[P_T^{-1}\right]$ and $\smax\left[P_T^{-1}\right]$
asymptotically scale linearly with~$T$. 
\par
\begin{figure}[tb]
\begin{center}
\includegraphics[width=0.47\textwidth]{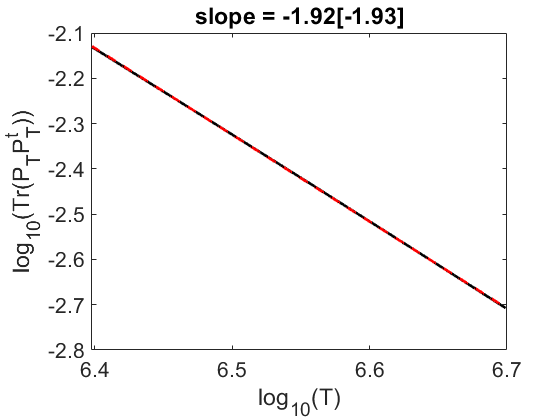}
\includegraphics[width=0.47\textwidth]{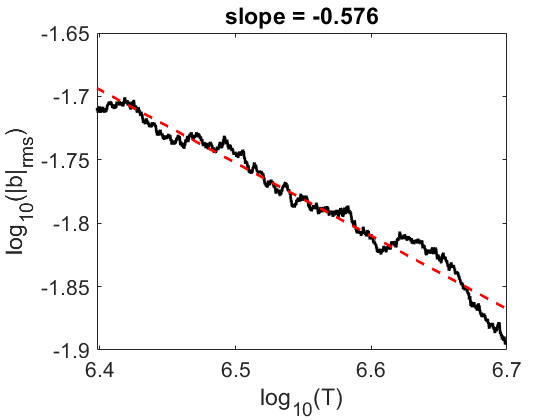}
\end{center}
\caption{\label{fig:longtime}Left: plotting the logarithm $\Tr\left(P_T^{\top}P_T\right)$
  versus the log of the number of iterations for $\sigma_w=0.1$\,mm shows
  that the slope is close to -2. Right: In the same way plotting the log
  of $|b_T|_{rms}$ shows a slope of approximately $\kappa=-0.57$. The red dashes
  denote the fitted straight lines.}
\end{figure}
For the asymptotic approach of the estimate $\hB_T$ towards the ``real'' response
matrix $B$ we insert the eigenvalues in Equation~\ref{eq:laiwei} and find
\begin{equation}\label{eq:Basymp}
  |\hB_T -B|_{\infty}
  = O\left(\sqrt{\frac{\log T}{T\smin\left[K\Gamma K^{\top}\right]}}\right)\ ,
\end{equation}
where we did not spell out constant factors.
In passing we note that $P_T = \left(U_T^{\top}U_T\right)^{-1}$ scales with
$1/\smin\left[P_T^{-1}\right]$ and this leads to
\begin{equation}\label{eq:PTasymp}
P_T=O\left(\frac{1}{\sigma_w^2T\smin\left[K\Gamma K^{\top}\right]}\right)\ ,
\end{equation}
which decreases like $1/T$ in the leading order.
\par
In order to verify the asymptotics numerically we run simulations with
$\sigma_w=0.1\,$mm for $5\times 10^6$ iterations. Figure~\ref{fig:longtime} shows
the asymptotic behavior of $\Tr\left(P_T^{\top}P_T\right)$ and of $|b_T|_{rms}$ as
a function of the iteration number on a double logarithmic scale in the range
between $2.5$ and $5\times 10^6$ iterations. A linear fit to the data on the
left-hand side shows a slope of $-1.92$, if fitting the entire range, and $-1.93$,
if fitting the upper 20\,\%. This indicates an approximate tendency towards
$\Tr{P_T^{\top}P_T}\propto 1/T^2$, which is consistent with Equation~\ref{eq:PTasymp}.
Repeating these calculations for different random seeds gives comparable results.
On the other hand, the slope of $|b_T|_{rms}$ is approximately $0.57$, which is
close to $1/\sqrt{T}$, the dominant dependence in Equation~\ref{eq:Basymp}.
But the the curve is much more noisy, which we attribute to the logarithm of $T$
in the numerator of Equation~\ref{eq:Basymp}.
\par
\begin{figure}[tb]
\begin{center} 
  \includegraphics[width=0.99\textwidth]{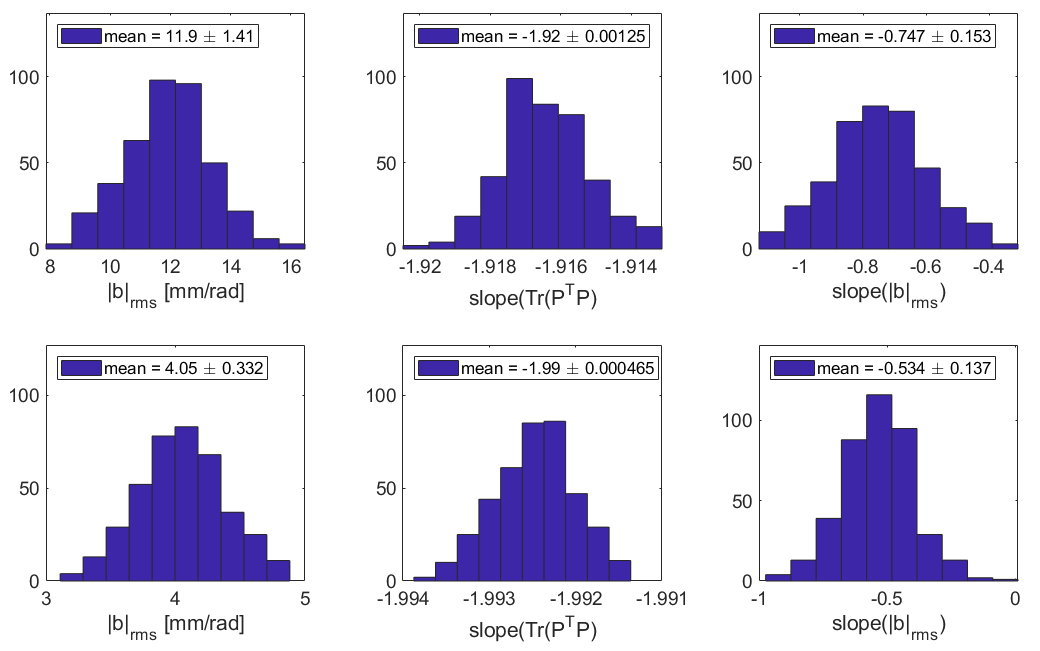}
\end{center}
\caption{\label{fig:verylong}Top row: histograms of the final discrepancy $|b_f|_{rms}$
  after $5\times10^6$ iterations (left) with $\sigma_w=0.1\,$mm, the slope of
  $\Tr\left(P_T^{\top}P_T\right)$ (center) and the slope of $|b_T|_{rms}$ (right)
  in the range $2.5\times 10^6$ to $5\times 10^6$ iterations. Bottom row: the
  corresponding plots with $20\,\mu$rad round-robin dither added.}
\end{figure}
In order to  explore this variability we run the simulation with 400 different
random seeds, all having $\sigma_w=0.1\,$mm, and plot the final value of the
discrepancy $|b_f|_{rms}$, the slope of  $\Tr\left(P_T^{\top}P_T\right)$, and the slope
of $|b_T|_{rms}$ in the top row of histograms in Figure~\ref{fig:verylong}. We see
that after $5\times 10^6$ iterations $|b_f|_{rms}$ has reached a value of about $12\,$mm/rad
(left). The slope of $\Tr\left(P_T^{\top}P_T\right)$ is $-1.92$ (center) and has not quite
reached its asymptotic value of $-2$. The asymptotic slope of $|b_T|_{rms}$ (right)
is approximately $0.75$. The width of the histograms indicate their
standard deviations, which is indicated as the uncertainty in the respective legends
of the plots. We observe that the results are reasonably stable and give a good indication of
the asymptotic behavior of the system. In the bottom row in Figure~\ref{fig:verylong}
we show the corresponding plots for the situation, where $20\,\mu$rad round-robin
dither is added. We find that the final value of $|b_f|_{rms}$ is only $4\,$mm/rad (left),
while the slope of $\Tr\left(P^{\top}P\right)$ is very close to the asymptotic value of $-2$.
The slope of the discrepancy $|b_T|_{rms}$ (right) indicates a value of
approximately $-0.53$. We point out that the width of the two histograms on the
right is much larger than the others, which we again attribute to the logarithm in
the numerator of Equation~\ref{eq:Basymp}.
\section{Some technical aspects}
\label{sec:tech}
We now turn to practical aspects of our system to determine the ``real'' response
matrix $B.$ From Equations~\ref{eq:Basymp} and~\ref{eq:PTasymp} we see that the most
important quantity for convergence is the smallest eigenvalue of $K\Gamma K^{\top}$,
where $\Gamma=\sum_{s=0}^{\infty} \left(\Lambda\Lambda^{\top}\right)^s$ is defined
immediately before Equation~\ref{eq:Euu}. For all well-behaved feedback systems
$\rho(\Lambda)=\rho(1-BK)$ is much smaller than unity and the term with $s=0$
dominates the sum, which makes $\Gamma$ very close to the $m\times m$ unit matrix.
Since we do not a priori know $B$, we just set $\Gamma$ to the unit matrix when
evaluating the performance of our system and consider $K K^{\top}$ alone.
\par
\begin{figure}[tb]
\begin{center}
\includegraphics[width=0.47\textwidth]{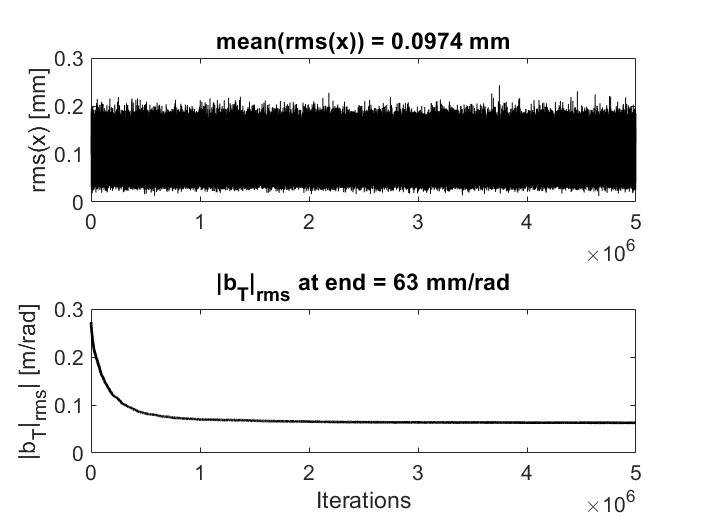}
\includegraphics[width=0.47\textwidth]{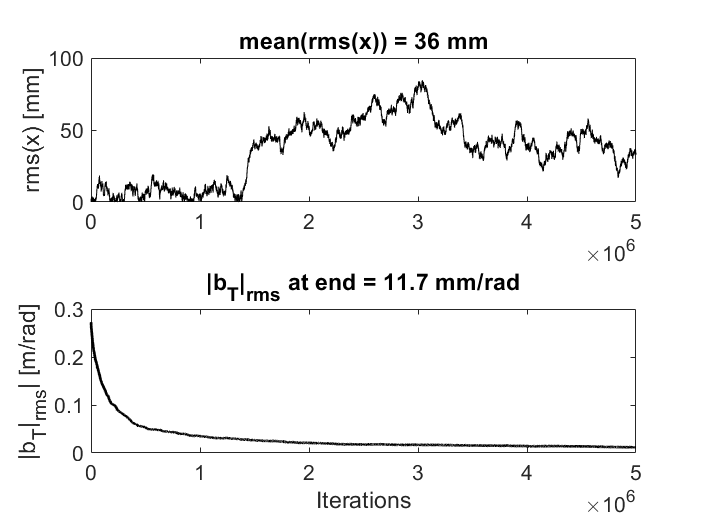}
\end{center}
\caption{\label{fig:oneoff}Left: one position monitor is removed ($n<m$),
  which spoils the convergence of $|b_T|_{rms}$ to zero, but maintains a
  small orbit. Right: one corrector magnet is removed ($n>m$); now the
  identification of the response matrix  works well and $|b_T|_{rms}$
  converges to zero, but the orbit is not corrected properly.} 
\end{figure}
If the feedback system is equipped with more correctors than position
monitors ($n<m$), the matrix $K K^{\top}$ is degenerate a has a null eigenvalue,
which spoils the convergence. The left-hand plot in Figure~\ref{fig:oneoff} shows what
happens when we remove one row, corresponding to one position monitor, from the response
matrix and repeat the analysis. The orbit, shown on the upper panel is still corrected
with a rms value comparable to $\sigma_w$, but $|b_T|_{rms}$, shown on the lower panel,
no longer converges to zero. The identification of the response matrix only works
partially and a finite difference to the ``real'' $B$ remains.
\par
If, on the other hand, there are more position monitors than corrector magnets
($n>m$)---in the simulation we removed one column, corresponding to one corrector
magnet, from the response matrix---the identification of the response matrix works
well, as illustrated on the lower panel on the right-hand plot in Figure~\ref{fig:oneoff},
because $m\times m$ matrix $K K^{\top}$ has full rank---no null eigenvalues. On the
other hand, we can no longer correct the orbit, as shown on the upper panel, because
now the $m\times n$ matrix $K$ now has eigenvalues null. We can, however, remedy this
problem by decomposing the symmetric $n\times n$ matrix $K^{\top}K=O D O^{\top}$,
where $D$ is a diagonal matrix containing the eigenvalues $d_i$ and $O$ is an
orthogonal matrix, whose columns are the corresponding eigenvectors $\ket{o_i}$.
We note that $\Phi=\sum_{i\ \mathrm{in\ nullspace}}\ket{o_i}\bra{o_i}$ is a projection matrix onto
the null-space of $K^{\top}K$, such that $\Psi=1-\Phi$ projects onto its orthogonal
complement, which is the space of BPM readings that the correctors can actually
affect. If we use  $\Psi\ket{x}$ instead of $\ket{x}$ when we apply the correction,
the null-modes never pile up and become unstable. If we apply this method to the
example from the right-hand side in Figure~\ref{fig:oneoff}, the orbit in the upper
panel looks very similar to the one on the left-hand plot. Since we always know $K$
(as opposed to $B$, which we do not know), we can always construct $\Psi$. Using the
projector $\Psi$ we can also use our algorithm if there are more BPM than correctors.
\par
For one-to-one orbit correction feedback systems with equal number of position
monitors and correctors ($n=m$) we just have to evaluate the eigenvalues of $KK^{\top}$
and possibly adjust $K$ by hand in order to speed up the convergence, albeit at the
expense of compromising the orbit correction to some extent. The details depend on
the particular accelerator and we will not dwell on this point further.
\par
  In order to understand the scaling of the convergence with system parameters, we consider
  rings with increasing number of $n=m$ cells with equal phase advance that contain one
  corrector and one BPM, each, which results in a near-circulant response matrix~\cite{MIRZA}.
  In numerical experiments we find
  that the largest eigenvalue of $B^{\top}B$ approximately increases with $n^2$. Since
  the correction matrix $K$ is normally close to the pseudo-inverse of $B$, we expect
  the smallest eigenvalue of $KK^{\top}$ to have an inverse dependence on $n^2$.
  Moreover, $B$ is proportional to a typical value of the beta function $\hat\beta$ in
  the ring, which makes $K\propto 1/\hat\beta$, such that we find
  $\smin\left[KK^{\top}\right]\propto\left(\sigma_w/n\hat\beta\right)^2$; the
  algorithm works best in small rings with noisy BPM.
\par
  It is instructive to compare the achievable error bars for the response matrix
  with those of an open loop measurement campaign, which are approximately given
  by $\sigma(B)_{o}\approx\sigma_w/\hat\theta$, where $\hat\theta$ is the amplitude
  of the corrector excitations. In Section~II we found that error bars of
  $\hat B$ from the closed-loop measurements are given by $\sigma(\hat B)^2
  =\diag\left[2\sigma_w^2(U_TU_T^{\top})^{-1}\right] = 2\sigma_w\diag\left[P_T\right]$. Moreover, during
  the early stages of the convergence, the eigenvalues of $P_T$ are given by
  Equation~13. We see that all eigenvalues decrease with $x_0/(1+x_0\lambda_iT)$,
  albeit at a slow time scale, characterized by the eigenvalues $\lambda_i$ of~$Q$.
  This process continues until the asymptotic regime is reached, as discussed
  near the start of Section~VI. In the  asymptotic regime $P_T$ continues to
  decrease as specified by Equation~26. We conclude that the error bars always
  get smaller and do so without limit. Additionally, Equation~25 implies that
  the approximation $\hat B$ asymptotically approaches the ``true'' response
  matrix $B$.
\par
  Finally,
  extending the algorithm to include settling time $\tau$, processing delay $d$, and 
relaxation into a new equilibrium with time scale $\tau_d$ is straightforward
by introducing unobservable state variables $\ket{\alpha_t}$ and $\ket{\beta_t}$.
Their dynamic behavior is described by
\begin{equation}\label{eq:upAB}
  \ket{\alpha_t}=\frac{\tau}{\tau+1}\ket{\alpha_{t-1}} + \frac{1}{\tau+1} \ket{u_{t-1-d}}
  \quad\mathrm{and}\quad
  \ket{\beta_{t}} = \frac{\tau_d}{\tau_d+1}\ket{\beta_{t-1}}+\frac{1}{\tau_d+1}B\ket{\alpha_t}
\end{equation}
with $\ket{u_{t-1-d}}=-K\ket{x_{t-1-d}}$. The delay $d$ and time constants $\tau$
and $\tau_d$ affect the stability of the closed-loop system, but we assume that
the feedback designer has chosen $K$ to ensure its stability. In the equation,
$\ket{\alpha_t}$ corresponds to the field inside the vacuum chamber that the
beam actually ``sees'' and $\ket{\beta_t}$, for example, the damping due to synchrotron
radiation. The observable beam position $\ket{x_t}$ then updates with $\ket{x_{t+1}}
=\ket{x_{t}}+\ket{\beta_{t}}$. We note that the left side of Equation~\ref{eq:upAB} enables
us to uniquely determine the $\ket{\alpha_t}$ from the $\ket{u_{t}}$, which makes them
quasi observable, provided $\tau$ and $d$ are known. Moreover, we find the
$\ket{\beta_{t}}=\ket{x_{t+1}}-\ket{x_{t}}$ from the $\ket{x_{t}}$, which turns the
right side of Equation~\ref{eq:upAB} into $(\tau_d+1)\ket{\beta_t} - \tau_d\ket{\beta_{t-1}}
=B\ket{\alpha_t}$. We observe that this equation has the same form as Equation~3
from the main text with one component of the left-hand side taking the place of
$x^i_{s+1}-x_s^i$ shifted by one time step. Likewise, $\ket{\alpha_t}$ takes the
place of $\ket{u_s}$. The analysis from the report up to Equations~\ref{eq:upP}
and~\ref{eq:upB} remains valid, but analyzing the convergence and the asymptotics
goes beyond the scope of the present report.
\section{Conclusions}
We applied standard system identification techniques, based on recursive least-squares
methods, to determine the response matrix in parallel to correcting the orbit in a storage
ring. Simulations show that the method works well, though it is rather slow and
requires a large number of iterations. The speed can, however, be increased significantly
by systematically adding small perturbations to the corrector magnets, so-called dithering.
In this way a small deterioration of the orbit quality can be balanced with the desire to
determine an accurate response matrix. We found that the convergence of $\hB_T$
to the ``real'' response matrix is governed by the eigenvalues $\lambda$ of the matrix
$Q$ from Equation~\ref{eq:Q} and we solved the time dependence of the discrepancy
$\hb_T=\hB_T-B$ with some approximations. We found in Equation~\ref{eq:bG} that $\hb_T$
scales with $1/T$, but only until the magnitude of $\hb_T$ becomes smaller than the
error bars of the fitting process, which scale with $1/\sqrt{T}.$ Once inside
the noise level, we found that the asymptotic behavior of the convergence has a
$\sqrt{\log(T)/T}$ dependence and is governed by the smallest eigenvalue of $KK^{\top}$.
In particular, both the error bars of the approximation $\hB_T$ and the difference between
$\hB_T$ and the ``real'' $B$ tend to zero in the limit of large $T$.
Furthermore, we found that those feedback systems with number of BPMs equal or larger
than the number of correctors ($n\geq m$) permit us to simultaneously stabilize the orbit
and to identify the response matrix $B$. 
\par
Several extensions of this work come to mind. First, optimizing the correction matrix $K$
such that the smallest eigenvalue of $K K^{\top}$ is as large as possible without
spoiling the orbit quality $\sigma_x$. Second, comparing different  correction strategies,
for example, deriving $K$ from ``optimal control'' quality measures that put a weight
both on the orbit $\sigma^2_x$ and the rms corrector excitation. Third,
finding an optimal strategy to make the dither amplitude $z$ time-varying, such that
global measure of performance that balances orbit correction and system identification
is minimized. The {\em regret}, studied for instance in~\cite{REGRET}, may serve as
an example. 
%
\section*{Acknowledements}
This work was supported in part by the Swedish Research Council (grant 2016-00861),
and the Swedish Foundation for Strategic Research (Project CLAS).
\appendix
\section{Sherman-Morrison formula}
\label{sec:appSM}
Here we show that $P_{T+1}$ is given by Equation~\ref{eq:upP} if its inverse $P_{T+1}^{-1}$
is given by $P_{T+1}^{-1}=P_T^{-1}+\ket{u_{T+1}}\bra{u_{T+1}}$. To show this, we explicitely
calculate $P_{T+1}^{-1}P_{T+1}$ and show that it evaluates to the unit matrix
\begin{eqnarray}
  P_{T+1}^{-1}P_{T+1}
  &=&\left[P_T^{-1}+\ket{u_{T+1}}\bra{u_{T+1}}\right]
      \left[P_T - \frac{P_T\ket{u_{T+1}}\bra{u_{T+1}}P_T}{1+\braket{u_{T+1}|P_T|u_{T+1}}}\right]\\
  &=&P_T^{-1}P_T + \ket{u_{T+1}}\bra{u_{T+1}} P_T\nonumber\\
  && -\frac{P_T^{-1}P_T \ket{u_{T+1}}\bra{u_{T+1}} P_T+\ket{u_{T+1}}\bra{u_{T+1}} P_T\ket{u_{T+1}}\bra{u_{T+1}} P_T}
     {1+\braket{u_{T+1}|P_T|u_{T+1}}}\nonumber\\
  &=&1+\ket{u_{T+1}}\bra{u_{T+1}} P_T
      -\frac{\ket{u_{T+1}}\left(1+\braket{u_{T+1}|P_T|u_{T+1}}\right)\bra{u_{T+1}} P_T}
      {1+\braket{u_{T+1}|P_T|u_{T+1}}}\nonumber\\
  &=& 1\ ,\nonumber
\end{eqnarray}
and we can use Equation~\ref{eq:upP} to update $P_T$ with the new information that
is encoded in the new corrector excitations $\ket u_{T+1}$. Note that $P_T=(U^{\top}_TU_T)^{-1}$
and its inverse are symmetric by construction for all $T$. This implies that the
order of multiplication of $P_{T+1}$ and its inverse does not matter and we also have
$P_{T+1}P_{T+1}^{-1}=1$.
\section{Response-matrix update}
\label{sec:appCC}
Here we follow~\cite{SYSINF2} and show that the update of the response matrix $\hB$
is accomplished by Equation~\ref{eq:upB}. We therefore write Equation~\ref{eq:defB} for
time step $T+1$
\begin{eqnarray}
  \ket{\hB^{i:}_{T+1}}
  &=&\left(U_{T+1}^{\top}U_{T+1}\right)^{-1}U_{T+1}^{\top}
  \left(\begin{array}{c} y^i_1\\ \vdots \\ y^i_{T+1}\end{array}\right)
  = P_{T+1}\left[\sum_{s=1}^T y^i_s\ket{u_s} +y^i_{T+1}\ket{u_{T+1}}\right]\nonumber\\
  &=&\left[P_T - \frac{P_T\ket{u_{T+1}}\bra{u_{T+1}}P_T}{1+\braket{u_{T+1}|P_T|u_{T+1}}}\right]
      \left[\sum_{s=1}^T y^i_s\ket{u_s} +y^i_{T+1}\ket{u_{T+1}}\right]\ .
\end{eqnarray}
Here we introduce the abbreviation $y^i_T=x^i_{T+1}-x^i_T$, exploit that
$U_{T+1}^{\top} =(\ket{u_1},$ $\dots, \ket{u_{T+1}})$, and finally express $P_{T+1}$
through Equation~\ref{eq:upP}. In the next step we multiply the two square brackets
and obtain four terms
\begin{eqnarray}
  \ket{\hB^{i:}_{T+1}}
  &=& P_T \sum_{s=1}^T y^i_s\ket{u_s} + P_T y^i_{T+1} \ket{u_{T+1}}
      - \frac{P_T\ket{u_{T+1}}\bra{u_{T+1}}P_T\sum_{s=1}^T y^i_s\ket{u_s}} {1+\braket{u_{T+1}|P_T|u_{T+1}}}
      \nonumber\\
  && \qquad -\frac{P_T\ket{u_{T+1}}\bra{u_{T+1}}P_T\ket{u_{T+1}}y^i_{T+1}} {1+\braket{u_{T+1}|P_T|u_{T+1}}}
  \nonumber\\
  &=&\ket{\hB^{i:}_{T}}+ P_T y^i_{T+1} \ket{u_{T+1}}
      - \frac{P_T\ket{u_{T+1}}\braket{u_{T+1}|\hB^{i:}_T}} {1+\braket{u_{T+1}|P_T|u_{T+1}}}\\
  &&\qquad -y^i_{T+1}\frac{P_T\ket{u_{T+1}}\bra{u_{T+1}}P_T\ket{u_{T+1}}} {1+\braket{u_{T+1}|P_T|u_{T+1}}} 
      \nonumber
\end{eqnarray}
where, according to Equation~\ref{eq:defB}, we identify the estimate in the previous
iteration~$T$ as $\ket{\hB^{i:}_{T}}=P_T\sum_{s=1}^T\ket{u_s} y^i_s$. Combining the second
and the fourth term, we arrive at
\begin{equation}
  \ket{\hB^{i:}_{T+1}} = \ket{\hB^{i:}_T}+\left[x^i_{T+2}-x^i_{T+1}-\braket{u_{T+1}|\hB^{i:}_T}\right]
  \frac{P_T\ket{u_{T+1}}}{1+\braket{u_{T+1}|P_T|u_{T+1}}}\ .
\end{equation}
Taking the transpose of this equation and stacking the rows on  top of each other then
leads to Equation~\ref{eq:upB}.
\section{Code for one iteration}
\label{sec:appCode}
The following function receives $\hB$ and $P$, as well as the recently measured
orbit $\ket{x}$ and the dither vector $\ket{z}$ as input and returns the updated
matrices $\hB_{new}$ and $P_{new}$ as well as the orbit $\ket{x_{new}}$ after the
correction is applied. Inside the function, first the externally defined noise level
$\sigma_w$, a constant correction matrix $\tilde B$, the ``real'' response matrix
$B$, and the correction matrix $K$ are supplied as global variables. Next,
using $K$, the new corrector values
$\ket{u}$ are calculated, dither $\ket{z}$ is added to the correctors, and the
new orbit $\ket{x_{new}}$ is calculated, including the noise $\ket{w}$, here implemented
as normally distributed random numbers. Then the auxiliary quantity $\bra{u_{T+1}}P_T$
is stored in the variable {\tt tmp} and the inverse of the denominator in the last
term in Equation~\ref{eq:upP} is calculated. The next two lines are straight
implementations of Equations~\ref{eq:upP} and~\ref{eq:upB}.
\begin{verbatim}
  function [Bhatnew,Pnew,xnew]=one_iteration4(Bhat,P,x,z)
  global sig Btilde Breal Bplus % noise, est., real, corr.
  %  u=-Bhat\x+z; % adaptive feedback
  u=-Bplus*x+z;                              % eq. 2 + dither
  xnew=x+Breal*u+sig*randn(size(x));         % eq. 2
  tmp=u'*P;                                  % <u|P
  denominv=1/(1+tmp*u);                      % 1/(1+<u|P|u>)
  Pnew=P-tmp'*tmp*denominv;                  % eq. 6
  Bhatnew=Bhat+(xnew-x-Bhat*u)*tmp*denominv; % eq. 7
\end{verbatim}
The figures in the main body of the report are produced by iterating this function.
Note that in the above code the correction matrix $K$ is fixed. We can,
however, easily make the feedback adaptive by simply replacing this line
in the code by \verb#u=-Bhat\x+z#, as indicated in the commented-out line.
In this way, always the most recent approximation for the matrix $\hB_T$ is
used when correcting the orbit.
\section{Spatially correlated monitor noise}
\label{sec:appCorr}
The assuption that the noise $\ket{w_t}$ of
position monitors is uncorrelated, is easily relaxed and in this appendix
we show spatially correlated noise, characterized by $\EE\left\{\ket{v_t}\bra{v_s}\right\}
=\sigma_w^2 \delta_{ts} C$ affects the rest of the results, where $\sigma_w^2C$
is the covariance matrix of the noise. Its matrix elements on the diagonal
$\sigma^2_wC_{ii}$ describe the square of the error bars of BPM $i$ and the
off-diagonal elements describe the correlations among different BPMs. Note
that we separated the magnitude of the noise ($\sigma_w^2$) from the correlations,
where $C$ is a positive definite and symmetric matrix with matrix elements of
order unity.
\par
Since $C$ is symmetric we can decompose it into orthogonal matrices $O$ and a
diagonal matrix. Since it is positive definite, all its eigenvalues are positive
and we can write the diagonal matrix as the square of another diagonal
matrix~$D$
\begin{equation}
C=OD^2O^{\top}\ .
\end{equation}
We will now use this representation of $C$ to transform the dynamical system
represented by Equation~\ref{eq:dynsys}, but with correlated noise $\ket{v_t}$
\begin{equation}\label{eq:dyncorr}
  \ket{x_{t+1}}=\ket{x_t}+B\ket{u_t} + \ket{v_t}\ .
\end{equation}
and multiply it with $D^{-1}O^{\top}$ from the left, which results in
\begin{equation}\label{eq:dyncorr}
  D^{-1}O^{\top}\ket{x_{t+1}}=D^{-1}O^{\top}\ket{x_t}+D^{-1}O^{\top}B\ket{u_t} + D^{-1}O^{\top}\ket{v_t}\ .
\end{equation}
With the transformed variables
\begin{equation}\label{eq:transf}
\ket{y_t} = D^{-1}O^{\top}\ket{x_t},\qquad B' = D^{-1}O^{\top}B,\qquad \ket{w_t} = D^{-1}O^{\top}\ket{v_t}\ .
\end{equation}
Equation~\ref{eq:dyncorr} reads
\begin{equation}
\ket{y_{t+1}} = \ket{y_{t}} + B'\ket{u_t} + \ket{w_t}\ ,
\end{equation}
where we have
\begin{equation}
\EE\left\{\ket{w_t}\bra{w_s}\right\}= D^{-1}O^{\top}\EE\left\{\ket{v_t}\bra{v_s}\right\}OD^{-1}
  = D^{-1}O^{\top}\sigma_w^2C\delta_{ts}OD^{-1}=\sigma_w^2\delta_{ts}\mathbf{1}\ .
\end{equation}
We find that this system is equivalent to the one from Equation~\ref{eq:dynsys},
such that we can directly use the methods developed in the main body of this report. We only
need to undo the transformation from Equation~\ref{eq:transf} in the end.
\par
If we apply this procedure to Equation~\ref{eq:upP} and~\ref{eq:upB} we find that these
equations are unchanged; the improvement of the $\hB$ does not depend on the noise as
long as there are perturbations. Only the changes of the controller $\ket{u_t}$ and the
resulting orbit changes matter.
\par
The correlation matrix $C$ does, however, affect the convergence of the algorithm. Using
correlated noise $\ket{v_t}$ instead of $\ket{w_t}$ in Equation~\ref{eq:EEuu}, we find that
its last equality becomes
\begin{equation}
  \EE\{\ket{u_T}\bra{u_T}\}
  =  \sigma_w^2K\sum_{s=0}^{T-1} \Lambda^sC\left(\Lambda^{\top}\right)^sK^{\top}+o(1)\ .
\end{equation}
Following the reasoning from the main body, the term with $s=0$ is dominant, which
gives us  $\EE\{\ket{u_T}\bra{u_T}\}=\sigma_w^2KCK^{\top}$ and the matrix $Q$ from
Equation~\ref{eq:Q} becomes $Q=\sigma_w^2KC K^{\top} + \frac{z^2}{m}\mathbf{1}$.
With this version of $Q$ the conclusions of Section~\ref{sec:convergence} remain
the same.
\par
Also the asymptotic behavior is affected by $C$. Equation~\ref{eq:asymG} becomes
\begin{equation}
\Gamma_T=\sum_{s=0}^{T-1} \Lambda^sC\left(\Lambda^{\top}\right)^s \leq \sum_{s=0}^{\infty} \rho(C)\rho(\Lambda)^{2s}\mathbf{1}
  = \frac{\rho(C)}{1-\rho(\Lambda)^2}\mathbf{1} < \infty
\end{equation}
where $\rho(C)$ is the spectral radius of $C$. This results in a slight redefinition
of $\Gamma=\lim_{T\to\infty}\Gamma_T$ which still is finite, which renders the remainder
of the Section~\ref{sec:conv} valid.
\bibliographystyle{plain}

%
%
\end{document}